\documentclass[pra,twocolumn,showpacs,superscriptaddress]{revtex4}
\usepackage{amsmath,amssymb,mathrsfs,bbm}
\usepackage[dvips]{graphicx}
\usepackage{hyperref}
\usepackage{paralist}

\begin{document}
\title{Geometric entanglement and quantum phase transitions\\ in two-dimensional quantum lattice models}

\author{Qian-Qian Shi}
\affiliation{Centre for Modern Physics, Chongqing University,
Chongqing 400044, China} \affiliation{College of Materials Science
and Engineering, Chongqing University, Chongqing 400044, China}

\author{Hong-Lei Wang}
 \affiliation{Laboratory of Forensic Medicine and Biomedical
 Information,
 Chongqing Medical University, Chongqing 400016, China}

\author{Sheng-Hao Li}
 \affiliation{Chongqing Institue of Engineering,
 Chongqing 400037, China}

\author{Sam~Young~Cho}
\altaffiliation{E-mail: sycho@cqu.edu.cn}
 \affiliation{Centre for Modern
Physics, Chongqing University, Chongqing 400044, China}
\affiliation{Department of Physics, Chongqing University, Chongqing
400044, China}

\author{Murray T. Batchelor}
 \affiliation{Centre for Modern
 Physics, Chongqing University, Chongqing 400044, China}
\affiliation{Mathematical Sciences Institute and Department of
 Theoretical Physics, Research School of Physics and Engineering,
 The Australian National University, Canberra ACT 2601, Australia}

\author{Huan-Qiang Zhou }
 \affiliation{Centre for Modern
 Physics, Chongqing University, Chongqing 400044, China}
 \affiliation{Department of Physics, Chongqing University, Chongqing
 400044, China}

\begin{abstract}
Geometric entanglement (GE), as a measure of multipartite entanglement,
has been investigated as a universal tool to detect phase transitions in quantum many-body lattice models.
In this paper we outline a systematic method to compute GE for two-dimensional (2D) quantum many-body lattice models
based on the translational invariant structure of infinite projected entangled pair state (iPEPS) representations.
By employing this method, the $q$-state quantum Potts model on the square lattice with $q \in \{2,3,4,5\}$
is investigated as a prototypical example. Further, we have explored three 2D Heisenberg models:
the antiferromagnetic spin-1/2 XXX and anisotropic XYX models in an external magnetic field,
and the antiferromagnetic spin-1  XXZ model. We find that continuous GE does not guarantee a
continuous phase transition across a phase transition point. We observe and thus classify three
different types of continuous GE across a phase transition point: (i) GE is continuous with maximum
value at the transition point and the phase transition is continuous, (ii) GE is continuous with maximum value
at the transition point but the phase transition is discontinuous, and (iii) GE is continuous with non-maximum
value at the transition point and the phase transition is continuous. For the models under consideration we
find that the second and the third types are related to a point of dual symmetry and a fully polarized phase, respectively.
\end{abstract}


\maketitle

\section{Introduction}

Of the various measures of entanglement in quantum many-body systems \cite{review},
geometric entanglement is a measure of the
multipartite entanglement contained in a pure state.
More precisely stated, the geometric entanglement (GE) quantifies the distance between a given
quantum state wavefunction and the closest separable (unentangled) state~\cite{review,hba,tcw}.
GE has been shown to serve as an alternative marker to locate
critical points for quantum many-body lattice systems undergoing
quantum phase transitions.
This was demonstrated explicitly for a number of one-dimensional (1D) quantum systems \cite{wei,ow}
for which the GE diverges near the critical point with an amplitude proportional to the central charge
of the underlying conformal field theory at criticality \cite{central,scopy}.
Moreover, for 1D quantum systems at criticality,
the leading finite-size correction to the GE per lattice site is universal \cite{qqs,jean,hu},
and related to the Affleck-Ludwig boundary entropy \cite{AL}.
For the Lipkin-Meshkov-Glick model, a system of mutually interacting spins embedded in a magnetic field for
which analytic results can be derived, the global GE of the ground state and the single-copy entanglement
behave as the entanglement entropy close to and at criticality~\cite{LMG}.

GE thus serves as a useful tool to investigate quantum criticality in
quantum many-body lattice systems.
Apart from some exceptions~\cite{hl,gehlw,thesis,roman,z3pott},
almost all work to date exploiting GE to study phase transitions
has been restricted to quantum systems in 1D.
This is mainly due to the difficulty to compute GE,
because it involves a formidable optimization over all possible separable states.
Indeed, the calculation of various entanglement measures has been shown recently to be NP-complete \cite{Ionnou,Huang}.
This is further compounded by the inherent computational difficulties posed by two-dimensional (2D) quantum systems.
Nevertheless, significant progress has been made to develop
efficient numerical algorithms to simulate 2D quantum many-body lattice
systems in the context of tensor network
representations~\cite{verstraete,vidal,gv2d,jwx,bvt,rosgv,pwe,sz,hco,wks,orusreview,tn_latest}.
The algorithms have been successfully exploited to compute, for example, the
ground-state fidelity per lattice site~\cite{zhou,zov,jhz,whl,lsh},
which has been established as a universal marker to detect quantum phase transitions
in many-body lattice systems.
Indeed, the ground state fidelity per lattice site is
closely related to the GE. Therefore, it is
natural to expect that there should be an efficient way to compute
the GE  in the context of tensor network algorithms.
This has been achieved for quantum many-body lattice systems with
periodic boundary conditions in one spatial
dimension in the context of the matrix product state representation~\cite{hu}.

Quantum phase transitions in 2D quantum lattice models can be investigated
using a number of different physical quantities, including local order parameters,
fidelity per lattice site, single-copy entropy and multi-partite entropy measurement, and GE
per lattice site.
As is well known, local order parameters are defined relating to the spontaneous symmetry breaking
of some symmetry group, which results in degeneracy of ground states.
Different degenerate ground states can be distinguished from different values of the local order parameters,
and continuous and discontinuous phase transitions can be identified from the continuous and discontinuous
behavior of the local order parameters.
Although demonstrated to be capable of detecting phase transitions, it is
not entirely clear if GE can distinguish different degenerate ground states and
identify both continuous and discontinuous quantum phase transitions.

To address this issue, we improve a systematic method \cite{gehlw} to efficiently
compute the GE per lattice site for 2D quantum
many-body lattice systems in the context of tensor network algorithms based on an infinite
projected entangled pair state (iPEPS) representation.
This method is used to evaluate the GE per lattice site for a number of distinct
models defined on the infinite square lattice.
These models are
(i) the quantum Ising model in a transverse field,
(ii) the $q$-state quantum Potts model with $q=3, 4$ and $5$,
(iii) the spin-$\frac12$ antiferromagnetic XXX model in an external magnetic field,
(iv) the spin-$\frac12$ antiferromagnetic XYX model in an external magnetic field,
and (v) the spin-$1$ XXZ model.
By comparing the behavior of the local order parameters and the GE per site for the different models,
it is seen that GE can detect both continuous and discontinuous quantum phase transitions.
However, we observe that the continuity of the GE per site does not necessarily match with the continuity of
local order parameters.

This paper is arranged as follows. The definition of GE is given in Section II.
Results using the GE per lattice site as a marker of quantum phase transitions in the
various 2D quantum models on the infinite square lattice are
given in Section III using the procedure outlined in Appendices A and B based on the iPEPS representation.
Discussion of the results and concluding remarks are given in Section IV.

\section{The geometric entanglement per lattice site}

For a pure quantum state $|\psi\rangle$ with $N$ parties, the GE,
as a global measure of the multipartite entanglement,
quantifies the deviation from the closest separable state $|\phi\rangle$.
For a spin system each party could be a single spin but could also be a block of contiguous spins.
The GE $E(|\psi\rangle)$ for an $N$-partite
quantum state $|\psi\rangle$ is expressed as~\cite{review,hba,tcw}
\begin{equation}
 E(|\psi\rangle)=-\log_2{\Lambda_{{\rm max}}^{2}}, \label{def}
\end{equation}
where $\Lambda_{\rm max}$ is the maximum fidelity between $|\psi\rangle$ and all possible
separable (unentangled) and normalized states $|\phi\rangle$, with
\begin{equation}
 \Lambda_{{\rm max} }={\rm \max_{|\phi\rangle }} \; |\langle\psi|\phi\rangle|.
\end{equation}
The GE per party ${\mathcal E}_{N}(|\psi\rangle)$ is then defined as
\begin{equation}
 {\mathcal E}_{N}(|\psi\rangle)=N^{-1}E(|\psi\rangle).
\end{equation}
It corresponds to the maximum fidelity per party $\lambda^{{\rm
max}}_{N}$, where
\begin{equation}
 \lambda^{{\rm max}}_{N}=\sqrt[N]{{\Lambda_{{\rm max}}}},
\end{equation}
or equivalently,
\begin{equation}
{\mathcal E}_{N}(|\psi\rangle)=-\log_2{{(\lambda^{{\rm
max}}_{N})}^{2}}.
\end{equation}
The relation (\ref{def}) is analogous to the relation between the free energy and the partition function.
Note that for unentangled states the GE is zero.

For our purpose, we shall consider a quantum many-body system on an
infinite-size square lattice, which undergoes a quantum phase transition
at a critical point in the thermodynamic limit. In this
situation, each lattice site constitutes a party, thus the GE
per party is the GE per lattice site, which is well defined in the thermodynamic limit
($N \rightarrow \infty$), since the contribution to the fidelity from
each party (site) is multiplicative.
In the infinite size limit we thus denote
\begin{equation}
{\mathcal E}=\lim_{N\to\infty} {\mathcal E}_{N}
\end{equation}
as the GE per site.

\section{geometric entanglement in infinite square lattice systems}

In this section we provide an analysis from the perspective of GE
of different 2D quantum spin systems undergoing different types of quantum phase transitions.
The GE per site has previously been applied to ground states of a number of
1D spin chains \cite{wei,ow,central,scopy,qqs,jean,hu,hl}
and some 2D models~\cite{hl,gehlw,thesis,roman,z3pott}
across primarily continuous phase transitions.
Our analysis here extends these previous studies to a wider range of 2D quantum
models and with more exotic situations.
Previously, the GE of ground states for the quantum Ising model in a transverse field and the XYX model on the square lattice
have been investigated with TRG~\cite{hl} and iPEPS~\cite{gehlw} approaches, both of which show a continuous behavior in the GE corresponding to a continuous phase transition.
Our results are based on the application and improvement of an iPEPS algorithm for the calculation
of the GE of 2D quantum systems \cite{gehlw}, as detailed in Appendices A and B.
Preliminary results along this direction were also reported in Ref.~\cite{thesis}.
The size of the truncation dimension $D$ controls the underlying accuracy of the iPEPS algorithm.
For the purposes of this study, which focusses on the general behavior of the GE in the vicinity of quantum critical points,
we consider sufficiently large values of $D$, compared to Refs.~\cite{gehlw,thesis}, to be confident of the observed behavior.
Refinements in the implementation of the basic iPEPS algorithm have been discussed recently \cite{tn_latest}.

\subsection{2D quantum Ising model in a transverse field}

We consider the hamiltonian
 \begin{equation}
  H = -\sum_{( \vec r, \vec r')}
     S^{[\vec r]}_x S^{[\vec r']}_x - \lambda \sum_{\vec r}  S^{[\vec r]}_z,
 \end{equation}
 where here and in later subsections $S^{[\vec r]}_{\alpha}(\alpha=x,y,z)$ are spin-$\frac12$ Pauli
 operators acting at site $\vec r$, with $( \vec r, \vec r')$ running over all possible nearest-neighbor
 pairs on the square lattice.
 The parameter $\lambda$ measures the strength of the transverse magnetic field,
 with a phase transition from a ferromagnetic phase with
 two degenerate ground states to a paramagnetic phase with a single ground state occurring at the
 critical value $\lambda_c$.
 This model has been widely studied via a number of different techniques.
 Of particular relevance here are the previous calculation of the GE using the
 tensor product state and tensor renormalization group approach~\cite{hl} and iPEPS \cite{gehlw,thesis}.

\begin{figure}[tb]
\includegraphics[width=0.6\linewidth]{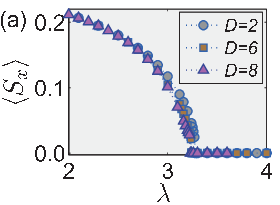}
\includegraphics[width=0.7\linewidth]{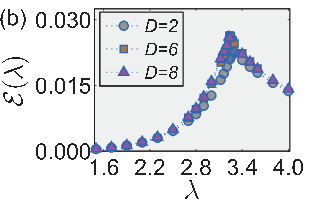}
\caption{(color online) Numerical values for (a) the local order parameter $\langle S_x\rangle$ and (b)
the GE per site ${\mathcal E}_{\infty}(\lambda)$ for the 2D quantum Ising model as a function of the
transverse field strength $\lambda$ for the indicated values of the iPEPS truncation dimension $D$.
   }
  \label{fig1}
\end{figure}

For different values of the truncation dimension $D$, when $\lambda<\lambda_c(D)$,
spontaneous symmetry breaking occurs with $S_x$ the unitary element of the broken $Z_2$ symmetry group.
Thus two degenerate ground states are detected and distinguished by the sign of the local
order parameter $\langle S_x\rangle$, with the amplitude associated with each of the two degenerate ground states
having the same value.
The order parameter $\langle S_x\rangle$ is shown in Fig.~\ref{fig1}(a).
Phase transition points are identified at the values shown in Table I.
Near these points the amplitude of $\langle S_x\rangle$ is observed to continuously approach zero,
indicative of a continuous phase transition.

On the other hand, the estimates for the GE per lattice site for the ground states are shown in
Fig.~\ref{fig1}(b).
For each value of $D$ the GE curve has a maximal value at $\lambda_c(D)$, which indicate the
phase transition points.
These values, shown in Table I, agree with the values detected with the local order parameter.
Our results are consistent with the previous studies \cite{hl,gehlw,thesis}.
The GE is seen to be continuous near the maximal points.
The cusp-like behavior at the critical point is a characteristic feature of the GE at a continuous phase transition.
The two degenerate ground states for $\lambda < \lambda_c$ can also be distinguished via the initial random states.

\begin{table}
\caption{\label{table} Estimates of critical points $\lambda_c$
for the $q$-state quantum Potts model on the square lattice.
The values for $\lambda_c(D)$ follow from the behavior of the
local order parameter and the GE per site,
where $D$ is the iPEPS truncation dimension.}
\begin{ruledtabular}
\begin{tabular}{lrlll}
~ & ~& $D$ & $\lambda_c(D)$  & \\
\hline
&$q=2$ & 2 & 3.28   &  \\
&~         & 6   & 3.235 &\\
&~         & 8   & 3.23  & \\
\hline
&$q=3$ & 3   & 2.620 & \\
&~         & 6   & 2.616 &\\
&~         & 9   & 2.616  &\\
\hline
&$q=4$ & 4   & 2.430 &\\
&~         & 8   & 2.428 &\\
&~         & 10 & 2.426 &\\
\hline
&$q=5$ & 5   & 2.330 &\\
&~         & 8   & 2.326 &\\
&~         & 10  & 2.326 &\\
\end{tabular}
\end{ruledtabular}
\end{table}


\subsection{2D $q$-state quantum Potts model}

The quantum Ising model discussed above is the special $q=2$ case of the
more general $q$-state quantum Potts model defined on the square lattice.
For a regular lattice, classical mean-field solutions~\cite{Wu}
and extensive computations (see, e.g., Refs.~\cite{Wu,dyw} and references therein)
have suggested that the 3D classical $q$-state Potts model, and thus the 2D $q$-state quantum version,
undergo a continuous phase transition for $q \le 2$ and a discontinuous phase transition for $q > 2$.
We now turn to this 2D $q$-state quantum model
and examine the GE per site and local order parameters in the vicinity of the
phase transition points for the values $q=3,4$ and $5$.

On the square lattice the hamiltonian can be written in the form
\begin{equation}
  H =
  -\sum_{( \vec r, \vec r')}
  \left( \sum_{p=1}^{q-1} M^{[\vec r]}_{x,p} M^{[\vec r']}_{x,q-p} \right)
    -\lambda \sum_{\vec r} M^{[\vec r]}_z,
\end{equation}
 where $M^{[\vec r]}_{x, p}$ and $M^{[\vec r]}_{z, p}$, with $p = 1,\ldots,q-1$,
 are spin matrices of size $q \times q$ acting at site $\vec r$.
 The parameter $\lambda$ is the analogue of the transverse magnetic field in the Ising case.
In this formulation the spin matrices acting at each site are given by \cite{qPotts}
\begin{equation*}
  M_{x,1} = \left(\begin{array}{cc} 0 & I_{q-1} \\ 1 & 0 \end{array} \right), \quad
  M_z = \left( \begin{array}{cc} q-1 & 0 \\ 0 & -I_{q-1} \end{array}
  \right),
\end{equation*}
 where
 $I_{q-1}$ is the $(q-1)\times (q-1)$ identity matrix with
 $M_{x,p} = \left(M_{x,1}\right)^p$ and $\left(M_{x,1}\right)^q = I_q$.

\begin{figure}[t]
\includegraphics[width=1.0\linewidth]{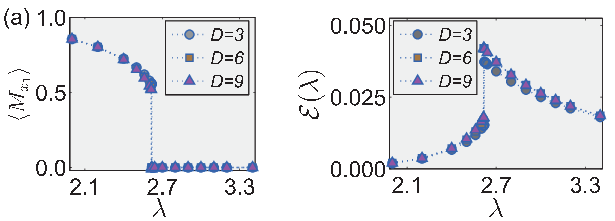}
\vskip 2mm
\includegraphics[width=1.0\linewidth]{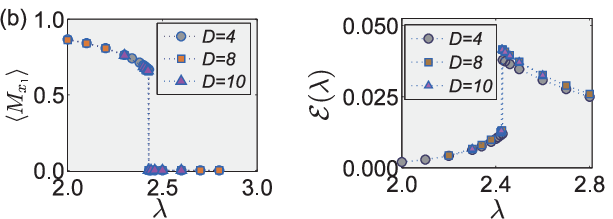}
\vskip 2mm
\includegraphics[width=1.0\linewidth]{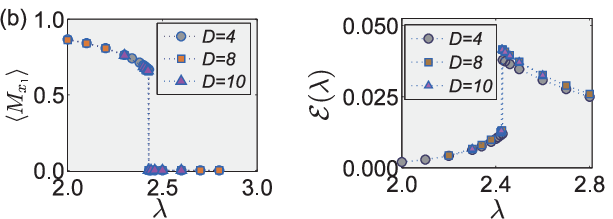}
\caption{(color online) Numerical values for the local order parameter $\langle M_{x_1}\rangle$ and
the GE per site ${\mathcal E}(\lambda)$ for the 2D quantum $q$-state Potts model
for (a) $q=3$, (b) $q=4$ and (c) $q=5$ as a function of the
transverse field strength $\lambda$ for the indicated values of the iPEPS truncation dimension $D$.
For each case the curves for both the local order parameter and the GE show a discontinuous behavior,
indicative of a discontinuous phase transition.
} \label{fig2}
\end{figure}

Denoting the phase transition point by $\lambda_c$ we expect to detect $q$-degenerate ground states for
$\lambda<\lambda_c$, corresponding to a $Z_q$ broken symmetry phase.
The local order parameter $\langle M_{x,1}\rangle$ can distinguish the different degenerate ground states,
but with the same amplitude for each of the $q$ ground states.
As for the $q=2$ case, the phase transition point $\lambda_c(D)$ is estimated
with increasing truncation dimension $D$.
Fig.~\ref{fig2}(a) shows the amplitude of the local order parameter $\langle M_{x,1}\rangle$ for the 3-state Potts model.
This plot shows a jump in the curve which indicates a possible discontinuous phase transition point.
The successive estimates for $\lambda_c$ are given in Table I \cite{ZN}.
This same type of discontinuous behavior is seen for the local order parameter $\langle M_{x,1}\rangle$
in Fig.~\ref{fig2}(b) for $q=4$ and Fig.~\ref{fig2}(c) for $q=5$.
The successive estimates for $\lambda_c$ are given in Table I.

The GE of the ground states is also shown in Fig.~\ref{fig2}.
For each of the values of $q$, a maximal value is detected for the GE curve, where a jump also occurs.
The estimates for the transition points $\lambda_c$ are
well matched with those obtained via the local order parameters and also via the observed
multi-bifurcation points in the magnetization~\cite{dyw}.
It is clear that the measure of GE can distinguish between discontinuous and continuous phase transitions in the 2D quantum
$q$-state Potts model.
To further test the utility of this approach we turn now to the investigation of GE in other 2D quantum models.


\subsection{2D spin-$\frac12$ XXX model in a magnetic field}

The spin-$\frac12$ antiferromagnetic XXX model on the square lattice has hamiltonian
\begin{equation}
  H =
  \sum_{( \vec r, \vec r')} \left( S^{[\vec r]}_x S^{[\vec r']}_x
    +  S^{[\vec r]}_y S^{[\vec r']}_y
    +  S^{[\vec r]}_z S^{[\vec r']}_z \right)-h \sum_{\vec r} S^{[\vec r]}_z,
\end{equation}
where $h$ is an external magnetic field along the $z$ direction.
This model has been studied via a barrage of different techniques~\cite{sandvik}.
Whereas magnetic order is normally ruled out in 1D Heisenberg models, this is
not the case for 2D Heisenberg models~\cite{mermin,sandvik}.
Thus in 2D the ground state can be non-magnetic, i.e., with a non-zero magnetization.
In the absence of the magnetic field the ground state of the Heisenberg model has antiferromagnetic (N\'{e}el) order
with a non-zero local staggered magnetization and infinitely degenerate ground states resulting from the breaking of
global SU$(2)$ spin rotation symmetry.
For $h\neq0$, infinitely degenerate ground states are detected resulting from the spontaneous symmetry breaking
of U$(1)$ symmetry in the $x$-$y$ plane.
As $h\rightarrow \infty$ it is anticipated the system will be fully polarized in the $z$ direction.
In fact this transition to the fully polarized state occurs at $h=4$.

\begin{figure}
\includegraphics[width=0.8\linewidth]{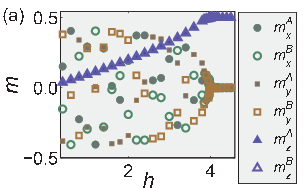}
\includegraphics[width=0.8\linewidth]{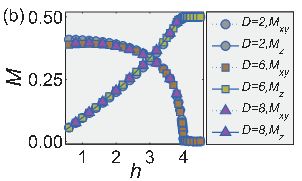}
\includegraphics[width=0.8\linewidth]{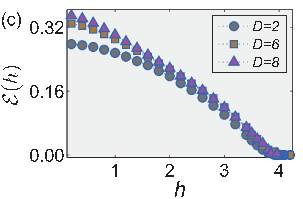}
\caption{(color online)
Plots for the 2D spin-$\frac12$ XXX model with varying magnetic field $h$.
(a) Local magnetization $m$ for sublattices A and B, obtained with iPEPS truncation dimension $D=2$.
(b) Amplitude of the local order parameters $M$.
(c) GE per site ${\mathcal E}(h)$.
  }
  \label{fig3}
\end{figure}

In Fig.~\ref{fig3}(a) we show the local magnetizations for sublattices A and B defined by
$m_{\alpha}^A=\langle S_{\alpha}^A\rangle$ and $m_{\alpha}^B=\langle S_{\alpha}^B\rangle$ for $\alpha=x,y,z$.
For $h<4$, $m_\alpha^A = -m_\alpha^B$ for $\alpha=x,y$ with $m_z^A=m_z^B$.
For $h>4$, $m_\alpha^A = m_\alpha^B = 0$ for $\alpha=x,y$ with $m_z^A=m_z^B =\frac12$.
The latter is indicative of the fully polarized state.

In Fig.~\ref{fig3}(b) we show the staggered magnetization parameters
$M_{xy}=\sqrt{\langle \frac12(S_x^A-S_x^B)\rangle^2+\langle \frac12(S_y^A-S_y^B)\rangle^2}$
and $M_z=\langle \frac12(S_z^A+S_z^B)\rangle$ as functions of the magnetic field.
The local order parameter $M_{xy}$ decreases with increasing $h$, with $M_{xy} \to 0$ when $h>4$.
On the other hand, $M_z$ increases with increasing $h$, with $M_{z} \to \frac12$ for $h>4$, which corresponds to
a continuous phase transition from a N\'eel phase to the fully polarized phase.
As mentioned previously, for $h<4$, infinitely degenerate ground states exist corresponding to the
$x$-$y$ plane U$(1)$ symmetry breaking, which is indicated from the random-like magnetization on the sublattices.
As for the previous models,  the numerical results indicate that the different degenerate ground states for
fixed $h$ give the same value for the GE per site, as Fig.~\ref{fig3}(c) shows.
In Fig.~\ref{fig3}(c), the GE per site clearly decreases with increasing $h$, indicating decreasing entanglement as $h\to 4$.
In fact, zero GE per site indicates a factorizing field, in this case the simple polarized state.
This factorizing field is discussed further below in the context of the more general $XYX$ model.

In contrast to the previous models studied here, the GE is not maximal at the phase transition point,
because the phase transition point is not the critical point which would correspond to maximum entanglement.
In this case zero or nonzero GE per site distinguishes between the two different phases in the model.

\subsection{2D spin-$\frac12$ XYX model in a magnetic field}

Tuning the anisotropy of the Heisenberg interactions leads to the more general
spin-$\frac12$ XYX antiferromagnetic model in a uniform magnetic field, defined on the square lattice by the hamiltonian
\begin{equation}
  H =
  \sum_{( \vec r, \vec r')} \left( S^{[\vec r]}_x S^{[\vec r']}_x
    +\Delta_{y}  S^{[\vec r]}_y S^{[\vec r']}_y
    +S^{[\vec r]}_z S^{[\vec r']}_z \right)
    -h \sum_{\vec r} S^{[\vec r]}_z.
\end{equation}
Varying the anisotropic exchange interaction parameter $\Delta_y$ leads to different behavior, with
the two cases $\Delta_{y}<1$ and $\Delta_{y}>1$ corresponding to an easy-plane and easy-axis behavior, respectively.
The quantum criticality of this model is well understood \cite{tros}, with an ordered phase below a critical field value $h_c$,
above which is a partially polarized state with field-induced magnetization reaching saturation as $h \to \infty$.
The ordered phase in the easy-plane (easy-axis) case arises by
spontaneous symmetry breaking along the $x$ ($y$) direction,
which corresponds to a finite value of the order parameter $M_x$ ($M_y$) below $h_c$.
At the transition point $h_c$, long-range correlations are destroyed.

Previous studies of this model
using GE \cite{gehlw,hl} focussed on the value $\Delta_{y}=0.25$ with the magnetic field as control parameter.
In this easy-plane region the model is known to undergo a
continuous quantum phase transition in the same universality class as the transverse Ising model \cite{tros}.
The GE was seen to have a cusp at the critical value $h_c$  \cite{gehlw,hl}
as observed for the transverse Ising model (recall Fig.~\ref{fig1}(b)).
Significantly, it is known that a factorizing field exists at the value $h_f = 2 \sqrt{2(1+\Delta_y)}$, with $h_f < h_c$,
where the ground state becomes a separable product state \cite{tros}.
At this point it follows that the GE vanishes, i.e., ${\mathcal E}_{\infty}(h_f) = 0$.
This was observed in the simulations using GE \cite{gehlw,hl}.
Other entanglement measures were also confirmed to vanish at this point \cite{tros,lsh}.

In this study we consider the fixed parameter value $h=0.25$ and vary the coupling $\Delta_{y}$.
The model is anticipated to undergo a transition at $\Delta_{y}=1$ from an antiferromagnetic phase
in the $x$ direction to an antiferromagnetic phase in the $y$ direction, with local order parameter the
staggered magnetization $M_x$ ($M_y$)
corresponding to the phase $\Delta_{y}<1$ ($\Delta_{y}>1$).

\begin{figure}
\includegraphics[width=0.75\linewidth]{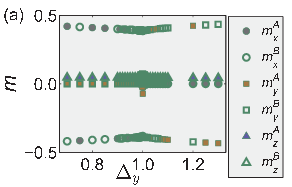}
\includegraphics[width=0.7\linewidth]{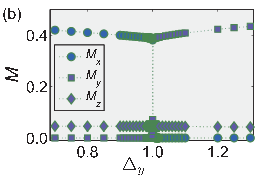}
\includegraphics[width=0.7\linewidth]{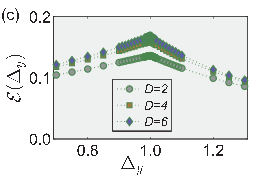}
\caption{(color online) Plots for the spin-$\frac12$ XYX model with fixed magnetic field $h=0.25$ and
varying anisotropy parameter $\Delta_y$.
(a) Components of the local magnetization $m$ for sublattices A and B for truncation dimension $D=6$.
 (b) Components of the staggered magnetizations $M$ for truncation dimension $D=6$.
 (c) GE per site ${\mathcal E}(\lambda)$ with truncation dimension $D$.}
\label{figXYX}
\end{figure}

Fig.~\ref{figXYX}(a) shows the components of the local magnetization for sublattices A and B defined by
$m_{\alpha}^A=\langle S_{\alpha}^A\rangle$ and $m_{\alpha}^B=\langle S_{\alpha}^B\rangle$ for $\alpha=x,y,z$.
For $\Delta_y<1$, $m_x^A$ and $m_x^B$ have opposite values, with $m_y^A=m_y^B=0$ and $m_z^A=m_z^B$.
This implies that a staggered magnetization $M_x=\langle \frac12 (S_x^A-S_x^B)\rangle$ exists.
For $\Delta_y>1$, $m_y^A$ and $m_y^B$ have opposite values, with $m_x^A=m_x^B=0$
and $m_z^A=m_z^B$, implying the staggered magnetization $M_y=\langle \frac12 (S_y^A-S_y^B)\rangle$.
These magnetizations are shown in Fig.~\ref{figXYX}(b) for the two different phases for truncation dimension $D=6$.
There is a clear jump discontinuity at $\Delta_y=1$ for each of the staggered magnetizations $M_x$ and $M_y$.
This behavior persists with increasing truncation dimension $D$, indicating a discontinuous phase transition at $\Delta_y=1$.

The GE per site is shown for the same parameter range in Fig.~\ref{figXYX}(c) with increasing truncation dimension.
The characteristic cusp occurs as $\Delta_{y}$ varies across the critical point $\Delta_{y}=1$.
In contrast to the local order parameter, which is discontinuous, the GE is continuous.
In this case the GE thus does not detect the discontinuous behavior at the phase transition point.


\subsection{2D spin-$1$ XXZ model}

\begin{figure}[t]
\includegraphics[width=0.7\linewidth]{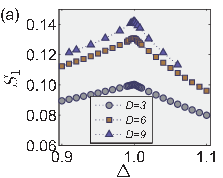}
\includegraphics[width=0.7\linewidth]{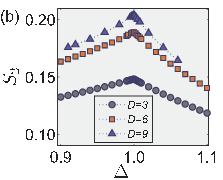}
\caption{(color online) Single-copy entanglement
  (a) $S_1(\Delta)$ and (b) $S_2(\Delta)$ as a function of the anisotropy parameter
  $\Delta$ for the 2D spin-$1$ XXZ model with increasing iPEPS truncation dimension $D$.
  Here the single-copy entanglement is calculated based on the one-site and two-site reduced density matrix,
  obtained by using the iTEBD method.}
\label{figXXZ_1}
\end{figure}

The spin-$1$ XXZ model is defined on the square lattice by the hamiltonian
\begin{equation}
  H =
  \sum_{( \vec r, \vec r')} \left( S^{[\vec r]}_x S^{[\vec r']}_x
    +  S^{[\vec r]}_y S^{[\vec r']}_y
    + \Delta S^{[\vec r]}_z S^{[\vec r']}_z \right),
\end{equation}
where now $S^{[\vec r]}_{\alpha} (\alpha=x,y,z)$ are spin-$1$ operators at site $\vec r$,
with again the summation running over all nearest neighbor pairs on the square lattice.
For anisotropic exchange interaction parameter $\Delta=1$ the hamiltonian has $SU(2)$ symmetry
and the ground state is infinitely degenerate.
For this model there is a quantum phase transition at $\Delta=1$ \cite{spin1}.
This phase transition is clearly marked in Fig.~\ref{figXXZ_1}, which shows plots of the single-copy
entanglement for the spin-$1$ XXZ model.

\begin{figure}[t]
\includegraphics[width=0.7\linewidth]{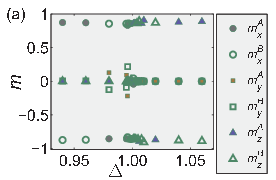}
\includegraphics[width=0.7\linewidth]{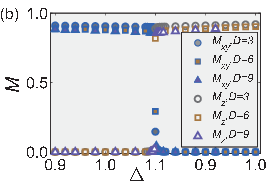}
\includegraphics[width=0.7\linewidth]{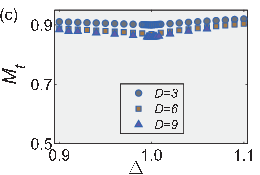}
\includegraphics[width=0.7\linewidth]{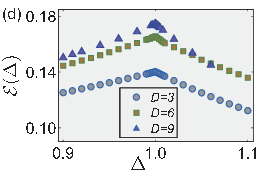}
\caption{(color online)
(a) Sublattice components of the local magnetization $m$,
(b) amplitude of the local order parameters $M_{xy}$ and $M_z$,
(c) total antiferromagnetic order parameter $M_t$ and (d) GE per lattice site
  ${\mathcal E}(\Delta)$ as a function of anisotropy parameter $\Delta$ for the 2D spin-$1$ XXZ model.
  Values of the iPEPS truncation dimension $D$ are as indicated, with $D=9$ for (a).}
\label{figXXZ_2}
\end{figure}

We also calculate the components of the local magnetization $m$ for sublattices $A$ and $B$:
$m_{\alpha}^A=\langle S_{\alpha}^A\rangle$ and $m_{\alpha}^B=\langle S_{\alpha}^B\rangle$ for $\alpha=x,y,z$.
These are shown in Fig.~\ref{figXXZ_2}(a) as a function of $\Delta$.
For $\Delta<1$, the values of $m_{x}$ and $m_{y}$ have opposite signs on each sublattice, with $m_z^A=m_z^B$.
For $\Delta>1$, $m_{x}=m_{y}=0$ on each sublattice,  with $m_z^A=-m_z^B$.
This indicates a local order with staggered magnetization in the $z$ phase characterized by
$
M_{z}=\langle \frac12 (S_z^A-S_z^B)\rangle.
$
For different values of magnetic field strength $h$, the magnetization for the $x$ and $y$ directions
does not change continuously for $\Delta<1$, while the staggered magnetization
$
M_{xy}=\sqrt{\langle \frac12(S_x^A-S_x^B)\rangle^2+\langle \frac12(S_y^A-S_y^B)\rangle^2},
$
continuously changes with varying $\Delta$ (see Fig.~\ref{figXXZ_2}(b)).
This indicates the existence of local order defined with the staggered magnetization in the $x$-$y$ plane.

Hence, for $\Delta<1$, the phase is characterized by the antiferromagnetic order parameter $M_{xy}$
in the $x$-$y$ easy plane.
Infinitely degenerate ground states exist corresponding to the $x$-$y$ plane $U(1)$ symmetry breaking, indicated
from the random-like magnetization of sublattices as shown in Fig.~\ref{figXXZ_2}(a).
For $\Delta>1$, the phase is characterized by the antiferromagnetic order parameter $M_z$ along the easy axis,
with doubly degenerate ground states corresponding to the spin-flip (or one-site translational invariant)
$Z_2$ symmetry breaking.
These order parameters are shown in Fig.~\ref{figXXZ_2}(b).
For each truncation dimension, a distinct jump is detected at $\Delta=1$, indicating the phase transition is possibly discontinuous.
The total antiferromagnetic order parameter
$
M_{t}=\sqrt{\langle \frac12 (S_x^A-S_x^B)\rangle^2+\langle \frac12 (S_y^A-S_y^B)\rangle^2
+\langle \frac12 (S_z^A-S_z^B)\rangle^2}
$
is shown in Fig.~\ref{figXXZ_2}(c).
It varies continuously.

For comparison we also consider the GE per site for this model.
The GE per site is seen to take the same value for all of the degenerate ground states.
As shown in Fig.~\ref{figXXZ_2}(d), the GE is continuous with a well defined cusp at the point $\Delta=1$,
while the local order parameters in Fig.~\ref{figXXZ_2}(b) are discontinuous at $\Delta=1$.
This behavior is similar to that observed in the previous subsection for the 2D spin-$\frac12$ XYX model with fixed
magnetic field $h$.

\section{summary and discussion}

We have demonstrated how to efficiently compute the GE of 2D quantum models defined on the
infinite square lattice, by optimizing over all possible separable states, in the context of the tensor network
algorithm based on the iPEPS representation.
Our results, in line with previous studies~\cite{hl,gehlw,thesis,roman,z3pott}, demonstrate that GE
is able to detect continuous quantum phase transitions and
factorized fields for different 2D quantum systems.
Continuous phase transitions are marked by a characteristic continuous cusp-like behavior of the GE per site at the critical point.
This behavior is evident in Fig.~\ref{fig1}(b) for the 2D quantum Ising model in a transverse field.
Additionally we have demonstrated that the GE can detect discontinuous phase transitions in the
$q$-state quantum Potts model for $q \ge 3$.
For the values of $q$ considered, a maximal value is detected for the GE curve, where the discontinuous transition occurs
(see Fig.~\ref{fig2}).
It is thus clear that the measure of GE can distinguish between continuous and discontinuous
phase transitions in the 2D quantum $q$-state Potts model.

This overall picture is not so simple, however.
It has been demonstrated recently that quantum phase transitions should be treated with caution,
at least with regard to the ground state entanglement spectrum \cite{CKS2014}.
In the present study, we have seen an example where the GE is continuous with non-maximum
value at the transition point and the phase transition is continuous.
This is the case for the spin-$\case12$ XXX model, for which the GE is not maximal at
the phase transition point $h=4$ at which the GE vanishes (see Fig.~\ref{fig3}(c)).
We have also seen examples where the GE per site shows continuous behavior at
discontinuous phase transition points.
For example, in the spin-$\case12$ XYX model and the spin-$1$ XXZ model,
the GE per site shows continuous behavior with well defined cusps at the transition point
(Fig.~\ref{figXYX}(c) and Fig.~\ref{figXXZ_2}(d)) but the corresponding
local staggered magnetizations (Fig.~\ref{figXYX}(b) and Fig.~\ref{figXXZ_2}(b)) are discontinuous.
We have also considered the single-copy entanglement $S_L$,
in which the system under consideration is divided into two parts,
one with $L$ lattice sites and the other with the other lattice sites.
It is known that the single-copy entanglement $S_L$ sets a bound for the GE~\cite{scopy},
i.e., ${\cal E}<S_L=-\log_2{\mu_L^1}$, with $\mu_L^1$ the largest eigenvalue of the
reduced density matrix $\rho_L$ for an $L$-site subsystem.
It is found that the single-copy entanglements $S_1$ and $S_2$ show the same continuous behavior
near the phase transition points for the models considered here,
as shown, e.g., in Fig.~\ref{figXXZ_1} for the spin-$1$ XXZ model.
It appears then for such systems, for which discontinuous phase transitions occur corresponding to different types
of symmetry breaking and the total magnetization is continuous, both multi-partite and bi-partite entanglement
measures are continuous.

This situation can be further understood as follows.
First, as already mentioned, due to the spontaneous symmetry breaking,
including the spontaneous breaking of continuous symmetry,
different degenerate ground states are detected in the broken symmetry phase
by different values of the local order parameter.
However, the GE takes the same value for all degenerate ground states.
In terms of entanglement, this can be understood from the GE being a global measure of entanglement,
which is the same for all degenerate ground states.
From a different perspective, we can suppose the closest separable state to the ground state
and the ground states share some property, i.e., we suppose the closest separable state of
different degenerate ground states breaks the same symmetry of the ground states
\footnote{
The GE by definition is given from the fidelity between two states.
However, in this case the two states are related,
i.e., the closet separable state to the ground state and the ground states share some property.
For two degenerate ground states $|\Psi_g^1\rangle$ and $|\Psi_g^2\rangle$ satisfying
$|\Psi_g^2\rangle=U|\Psi_g^1\rangle$, the closest separable state
of different degenerate groundstate $|\phi^1\rangle$  and  $|\phi^2\rangle$
has the same relationship, i.e.,  the two closest separable states to the two states would correspondingly
satisfy $|\phi^2\rangle=U|\phi^1\rangle$.
Then the fidelity between the groundstate $|\Psi_g^1\rangle$ and its closest separable state
$|\phi^1\rangle$ can be written as
$\langle \Psi_g^1 |\phi^1\rangle=\langle \Psi_g^1 |UU^{\dag}|\phi^1\rangle=\langle \Psi_g^2 |\phi^2\rangle$,
which thus gives the same value for the GE.}.

Also from the perspective of symmetry, it could be argued for the
spin-$\case12$ XYX model and the spin-$1$ XXZ model
that the discontinuity due to the isometry cannot be
detected with the GE measure because the hamiltonian/ground state obeys a dual symmetry
either side of the phase transition point.
For the spin-$\case12$ XYX model this symmetry is in $\Delta_y$, with $\Delta_y=1$ the ``self-dual" point
\footnote{Supposing the coupling in front of the two-body interaction term
$S_z^r S_z^{r'}$ is $\Delta_z$, with $\Delta_z=1$ in the XYX model,
the Hamiltonian obeys a duality relation under the transformation
$\sigma_x \leftrightarrow\sigma_y$, $\Delta_y \rightarrow 1/\Delta_y$,
$\Delta_z \rightarrow\Delta_z/\Delta_y$, $h \rightarrow h/\Delta_y$.}.
Given such a duality of the hamiltonian, both the ground state and its closest separable state are also each dual under this transformation,
so the fidelity between the ground state and its closest separable state also obeys the same relation,
leading to continuous behavior across the phase transition point.
Similarly for the spin-$1$ XXZ model, the hamiltonian is symmetric at $\Delta=1$, with $\Delta \neq 1$ breaking the
symmetry into easy-plane and easy-axis.
In this case such a duality in the total staggered magnetization of the easy-plane and easy-axis would imply a duality of
a local property of the ground state wavefunction, leading to continuity in the GE.

We have seen then three different types of continuous GE across a phase transition point:

\vskip 1mm
\noindent (i) GE is continuous with maximum value at the transition point and the phase transition is continuous,

\vskip 1mm
\noindent (ii) GE is continuous with maximum value at the transition point but the phase transition is discontinuous, and

\vskip 1mm
\noindent (iii) GE is continuous with non-maximum value at the transition point and the phase transition is continuous.

\vskip 1mm
\noindent
For the models under consideration the second and third types are related to a point of dual
symmetry and a fully polarized phase, respectively.
Given this refinement in our understanding of GE as a marker of quantum phase transitions,
and the development of powerful tensor network algorithms, we can be confident that GE can be used as an alternative
route to explore quantum criticality in quantum lattice models.

\vskip 5mm

{\it Acknowledgements.}
 MTB gratefully acknowledges support from Chongqing University and
 the 1000 Talents Program of China.
 This work is supported in part by the
 National Natural Science Foundation of China (Grant Numbers 11575037, 11374379 and 11174375).

 \appendix
  \renewcommand{\appendixname}{Appendix~}

\section{The infinite projected entangled pair state algorithm}

Our aim is to compute the GE per lattice site for a
quantum many-body lattice system on an infinite-size square lattice
in the context of the iPEPS algorithm~\cite{gv2d}.
Here we follow the presentation given in Ref.~\cite{gehlw}.

\begin{figure}[t]
\includegraphics[width=0.48\textwidth]{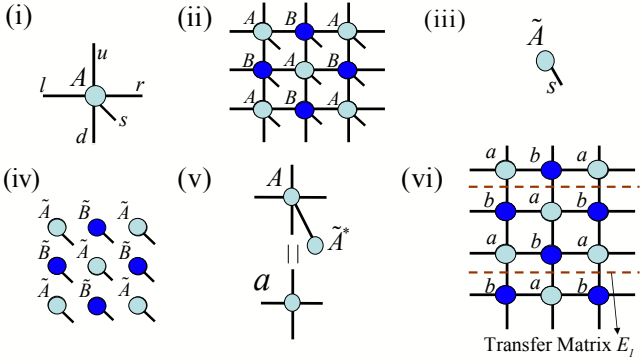}
\caption{(color online) (i) A five-index tensor $A^s_{lrud}$ labeled
by one physical index $s$ and four bond indices $l$, $r$, $u$ and
$d$. (ii) The iPEPS representation of a wavefunction on the
square lattice. Copies of the tensors $A^s_{lrud}$
and $B^s_{lrud}$ are connected through four types of bonds. (iii) A
one-index tensor $\tilde{A}^{s}$ labeled by one physical index $s$.
(iv) The iPEPS representation of a separable state in the
square lattice. (v) A reduced four-index tensor
$a_{lrud}$ from a five-index tensor $A^s_{lrud}$ and a one-index
$\tilde{A}^{s*}$.  (vi) The tensor network representation for the
fidelity between a quantum wavefunction (described by $A^s_{lrud}$
and $B^s_{lrud}$) and a separable state (described by $\tilde{A}^s$
and $\tilde{B}^s$), consisting of the reduced tensors $a_{lrud}$ and
$b_{lrud}$. } \label{FIG1}
\end{figure}

Suppose we consider a system characterized by a translation-invariant
Hamiltonian $H$ with nearest-neighbor interactions:
$H=\sum_{\langle ij \rangle} h_{\langle ij \rangle}$, with $h_{\langle ij \rangle}$ being the Hamiltonian density.
Assume that a quantum wavefunction $|\psi\rangle$ is
translation-invariant under two-site shifts, then one only needs two
five-index tensors $A^s_{lrud}$ and $B^s_{lrud}$ to express the
iPEPS representation.
Here, each tensor is labeled by one physical
index $s$ and four bond indices $l$, $r$, $u$ and $d$, as shown in Fig.~\ref{FIG1}(i).
Note that the physical index $s$ runs over
$1,\ldots,\mathrm{d}$, and each bond index takes $1,\ldots,D$, with
$\mathrm{d}$ being the physical dimension, and $D$  the bond
dimension. Therefore, it is convenient to choose a $2\times 2$
plaquette as the unit cell (cf. Fig.~\ref{FIG1}(ii)).
The ground state wavefunction is well approximated by
$|\psi_{\tau}\rangle$, which is obtained by performing an imaginary
time evolution~\cite{gv2d} from an initial state $|\psi_{0}\rangle$,
with
$|\psi_{\tau}\rangle=e^{-H\tau}|\psi_{0}\rangle/||e^{-H\tau}|\psi_{0}\rangle||$~\cite{gv2d},
as long as $\tau$ is large enough.

A key ingredient of the iPEPS algorithm is to take advantage of the
Trotter-Suzuki decomposition that allows to reduce the (imaginary)
time evolution operator $e^{-H \delta\tau}$ over a time slice
$\delta \tau$ into the product of a series of two-site operators,
where the imaginary time interval $\tau$ is divided into $M$ slices:
$\tau = M \delta \tau$.
Therefore, the original global optimization problem becomes a local two-site optimization problem.
With an efficient contraction scheme available to compute the effective
environment for a pair of the tensors $A^s_{lrud}$ and $B^s_{lrud}$~\cite{gv2d},
one is able to update the tensors $A^s_{lrud}$ and
$B^s_{lrud}$. Performing this procedure until the energy per lattice
site converges, the ground state wavefunction is produced in the
iPEPS representation.

\begin{figure}[t]
\includegraphics[width=0.48\textwidth]{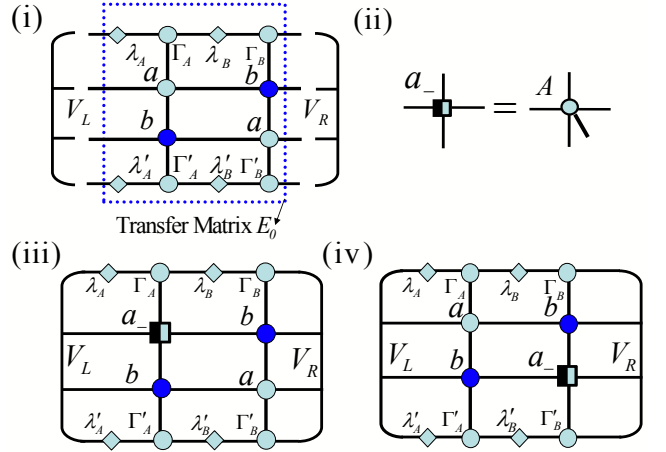}
\caption{(color online) Key ingredients for obtaining the
gradient of the fidelity between a given
ground state wavefunction $|\psi \rangle$ and a separable state
$|\phi \rangle$ in the iPEPS representation. (i) Zero-dimensional transfer matrix $E_0$ and its dominant eigenvectors
$V_{L}$ and $V_{R}$. Here the infinite matrix product state
representation of the dominant eigenvectors for  the one-dimensional
transfer matrix $E_1$ follows from Ref.~\cite{rosgv}, and $V_{L}$
and $V_{R}$ may be evaluated using the Lanczos method. The
contraction of the entire tensor network is the dominant eigenvalue
$\eta_{\langle\phi|\psi\rangle}$ of the zero-dimensional transfer
matrix $E_0$ for $\langle\phi|\psi\rangle$. (ii) A half-filled
square denotes $a_{-}$, the derivative of the four-index tensor
$a_{lrud}$ with respect to $\tilde{A}^{s*}$, which is nothing but
the five-index tensor $A^s_{lrud}$. Similarly, we may define
$b_{-}$, the derivative of the four-index tensor $b_{lrud}$ with
respect to $\tilde{B}^{s*}$. (iii) and (iv) Pictorial
representation of the contributions to the derivative of
$\eta_{\langle\phi|\psi\rangle}$ with respect to $\tilde{A}^{s*}$,
with different relative positions between filled circles and
half-filled squares. } \label{FIG2}
\end{figure}

\begin{figure}[th]
\includegraphics[width=0.48\textwidth]{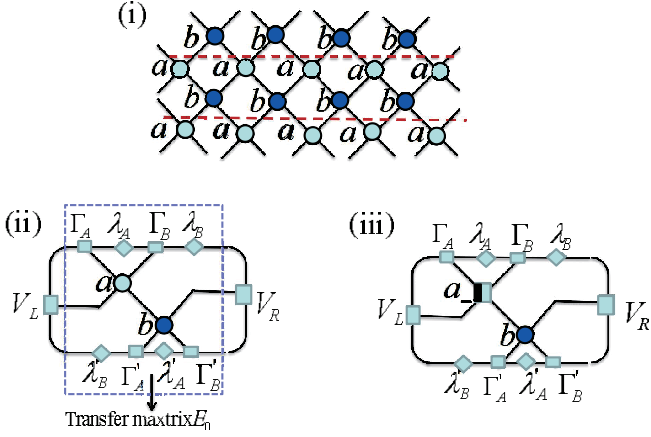}
\caption{(color online) Key ingredients in the diagonal contraction scheme for obtaining the
gradient of the fidelity between a given
ground state wavefunction $|\psi \rangle$ and a separable state
$|\phi \rangle$ in the iPEPS representation. (i) Tensor network representation for the
fidelity between a quantum wavefunction $|\psi \rangle$ and a separable state
$|\phi \rangle$.
(ii) Tensor network representation for a
zero-dimensional transfer matrix $E_0$ and its dominant eigenvectors
$V_{L}$ and $V_{R}$. Here the infinite matrix product state
representation of the dominant eigenvectors for the one-dimensional
transfer matrix $E_1$ follows from Ref.~\cite{rosgv}, and $V_{L}$
and $V_{R}$ are evaluated using the Lanczos method. The
contraction of the entire tensor network is the dominant eigenvalue
$\eta_{\langle\phi|\psi\rangle}$ of the zero-dimensional transfer
matrix $E_0$ for $\langle\phi|\psi\rangle$. (iii) The pictorial
representation of the contributions to the derivative of
$\eta_{\langle\phi|\psi\rangle}$ with respect to $\tilde{A}^{s*}$,
with different relative positions between filled circles and
half-filled squares. } \label{diagnal}
\end{figure}

\section{Efficient computation of the GE in the
iPEPS representation}

Once the iPEPS representation for the
ground state wavefunction is generated, we are ready to evaluate
the GE per lattice site.
Here we begin by outlining the scheme developed in Ref.~\cite{gehlw}.

First, we need to compute the fidelity between the ground state wavefunction and a
separable state. The latter is represented in terms of one-index
tensors $\tilde A^s$ and $\tilde B^s$. To this end, we form a
reduced four-index tensor $a_{lrud}$ from the five-index tensor
$A^s_{lrud}$ and a one-index tensor $\tilde A^s$, as depicted in
Fig.~\ref{FIG1}(iii). As such, the fidelity is represented as a
tensor network in terms of the reduced tensors $a_{lrud}$ and
$b_{lrud}$ (cf. Fig.~\ref{FIG1}(iv)).

The tensor network may be
contracted as follows. First, form the 1D transfer
matrix $E_1$, consisting of two consecutive rows of the tensors in
the checkerboard tensor network. This is highlighted in
Fig.~\ref{FIG1}(vi) by the two dashed lines. Second, compute the
dominant eigenvectors of the transfer matrix $E_1$, corresponding to
the dominant eigenvalue. This can be done, following a procedure
described in Ref.~\cite{rosgv}. Here, the dominant eigenvectors are
represented in the infinite matrix product states.  Third, choose
the zero-dimensional transfer matrix $E_0$ (Fig.~\ref{FIG2}(ii)), and compute its dominant
left and right eigenvectors, $V_{L}$ and $V_{R}$. This may be
achieved by means of the Lanczos method. In addition, one also needs
to compute the norms of the ground state wavefunction
$|\psi\rangle$ and a separable state $|\phi\rangle$ from their iPEPS
representations. Putting everything together, we are able to obtain the
fidelity per unit cell between the ground state $|\psi\rangle$ and a
separable state $|\phi\rangle$:
\begin{equation}
  \lambda=\frac{|\eta_{\langle\phi|\psi\rangle}|}
  {\sqrt{\eta_{\langle\psi|\psi\rangle}\eta_{\langle\phi|\phi\rangle}}},
\end{equation}
where  $\eta_{\langle\phi|\psi\rangle}$, $\eta_{\langle\phi|\phi\rangle}$ and
$\eta_{\langle\phi|\phi\rangle}$ are, respectively, the dominant
eigenvalue of the zero-dimensional transfer matrix $E_0$ for the
iPEPS representation of $\langle\phi|\psi\rangle$, $\langle\psi|\psi\rangle$ and
$\langle\phi|\phi\rangle$.

We then proceed to compute the GE per
site, which involves the optimization over all the separable states.
For our purpose, we define $F=\lambda^{2}$. The optimization amounts
to computing the logarithmic derivative of $F$ with respect to
$\tilde{A}^{*}$, which is expressed as
\begin{equation}
G \equiv \frac{\partial \ln F}{\partial
\tilde{A}^{*}}=\frac{1}{\eta_{\langle\phi|\psi\rangle}}\frac{\partial\eta_{\langle\phi|\psi\rangle}}{\partial
\tilde{A}^{*}}-\frac{1}{\eta_{\langle\phi|\phi\rangle}}\frac{\partial\eta_{\langle\phi|\phi\rangle}}{\partial
\tilde{A}^{*}}.
\end{equation}

The problem therefore reduces to the computation of $G$ in the
context of the tensor network representation. First, note that a
pictorial representation of the derivative $\partial a_{lrud}/
\partial \tilde{A}^{s*}$ of the four-index tensor $a_{lrud}$ with respect to
$\tilde{A}^{s*}$ is shown in Fig.~\ref{FIG2}(ii), which is nothing
but the five-index tensor $A^s_{lrud}$. Similarly, we may define the
derivative of the four-index tensor $b_{lrud}$ with respect to
$\tilde{B}^{s*}$. Then, we are able to represent the contributions
to the derivative of $\eta_{\langle\phi|\psi\rangle}$ with respect
to $\tilde{A}^{s*}$ in Fig.~\ref{FIG2}(iii) and Fig.~\ref{FIG2}(iv). In our
scheme, we update the real and imaginary parts of $\tilde{A}^s$
separately:
\begin{eqnarray*}
\Re(\tilde{A}^s)&=&\Re(\tilde{A}^s)+\delta \Re(G)^{s},\\
\Im(\tilde{A}^s)&=&\Im(\tilde{A}^s)+\delta \Im(G)^{s}.
\end{eqnarray*}
Here $\delta\in[0,1)$ is the step size in the parameter space,
which is tuned to be decreasing during the optimization process.
In addition, we have normalized the real and imaginary parts of the
gradient $G$ so that their respective largest entries are unity.
The procedure to update the tensor $\tilde{B}^s$ is the same.
If the fidelity per unit cell converges, then the closest separable state
$|\phi\rangle$ is achieved, thus the geometric entanglement per
lattice site for the ground state wavefunction $|\psi\rangle$ follows.

The above scheme has some basic limitations for the practical calculation of the GE per site.
In particular, it can only be readily achieved for relatively small truncation dimensions.
Larger truncation dimensions can be achieved by changing the transfer matrix direction in
Fig.~\ref{FIG1}(vi) to the diagonal direction, as shown in Fig.~\ref{diagnal}(i).
This leads to a simpler calculation of the derivatives necessary for the calculation of the GE
(see Fig.~\ref{diagnal}(ii) and Fig.~\ref{diagnal}(iii)).
For both schemes, the leading computational time scale during contraction is the same, namely $O(D^4)$.
However, in the diagonal scheme,
the computation time scale relative to truncation dimension $D$ can be reduced.
For example, to calculate the largest eigenvalue $\eta_{\langle\phi|\psi\rangle}$ of $E_0$
the computation time for each Lanczos step can be reduced from
$O(D^4)$ to $O(D^2)$ by storing tensors whose computation time scales as $O(D^4)$.
Since the time cost to obtain $\eta_{\langle\phi|\psi\rangle}$ is mainly due to iterations in the Lanczos algorithm,
this improvement allows us to deal with relatively larger truncation dimensions and thus
obtain higher accuracy for the GE per site.


\end{document}